\documentstyle[aps,prl,multicol,epsf]{revtex}
\newcommand{\beq}{\begin{equation}}
\newcommand{\eeq}{\end{equation}}
\newcommand{\beqa}{\begin{eqnarray}}
\newcommand{\eeqa}{\end{eqnarray}}

\begin{document}

\title{\bf Interface Fluctuations, Burgers Equations, and Coarsening 
under Shear}

\author{Alan J. Bray, Andrea Cavagna, and Rui D. M. Travasso}

\address{
Department of Physics and Astronomy, University of Manchester,  
Manchester, M13 9PL, UK}

\date{\today}

\maketitle

\begin{abstract}

We consider  the interplay  of thermal fluctuations  and shear  on the
surface of the domains in  various systems coarsening under an imposed
shear  flow.  These  include systems  with nonconserved  and conserved
dynamics, and  a conserved order  parameter advected by a  fluid whose
velocity field satisfies the Navier-Stokes equation.  In each case the
equation of motion for the  interface height reduces to an anisotropic
Burgers equation.  The scaling exponents that describe  the growth and
coarsening of the interface are calculated exactly in any dimension in
the case of conserved and nonconserved dynamics.  For a fluid-advected
conserved  order parameter  we  determine the  exponents,  but we  are
unable to  build a consistent perturbative expansion  to support their
validity.

\noindent

\end{abstract}

\begin{multicols}{2}

\section{Introduction}

This  paper   deals  with  the  influence  of   shear  on  interfacial
fluctuations  in  phase-ordering  or  phase-separating  systems.   The
primary motivation is the need  to understand the influence of thermal
fluctuations on  coarsening under shear. Thermal  fluctuations are not
normally  thought  to be  important  for  coarsening  systems, as  the
dynamics   is    controlled   by   a    ``strong   coupling'',   i.e.\
zero-temperature,   fixed  point  and   temperature  is   formally  an
irrelevant  perturbation \cite{Review}.   Under an  externally imposed
shear flow, however, the growing  domains become stretched in the flow
direction  \cite{experiment,theory,beppe,padilla,orange,shou,berthier}
and  there is  evidence, especially  in two  spatial  dimensions, that
growth   in   the   transverse   direction  is   strongly   suppressed
\cite{padilla,orange,shou,berthier}.  This raises the possibility that
thermal  roughening  of the  interface  might  destroy the  coarsening
state. On the other hand,  the thermal roughening is itself suppressed
by the shear  flow, so the question of the  survival of the coarsening
regime to  late times  rests on a  delicate balance between  these two
effects.

A  second motivation  for  this study  emerges  from the  mathematical
description of  the interfacial fluctuations, which takes  the form an
anisotropic Burgers equation \cite{burgers,fns}. The structure  of the
equation, and  the form of the  noise correlator, are such  that, in a
renormalization group (RG) analysis, some parameters of the theory are
not perturbatively renormalized.  As a result, certain combinations of
scaling exponents can be determined exactly. Remarkably, the number of
such  combinations is in  every case  equal to  the number  of unknown
exponents, so that all scaling exponents can be determined exactly for
any spatial dimensionality $d$.

The structure  of the interface  equation is very simple.   If $h({\bf
x},t)$ is  the interfacial height  relative to the mean  height, where
${\bf x}$  is a $(d-1)$-dimensional vector specifying  position in the
plane  parallel to  the (mean)  interface, and  $t$ is  the  time, the
equation takes the simple form
\begin{equation}
\partial_t h + \gamma h\partial_x h  = {\cal L}\,h + \eta({\bf x},t) \
,
\label{burgers}
\end{equation}
where $\gamma$ is the shear rate,  and we have taken the shear flow to
be in the $x$ direction.  The linear operator $\cal{L}$ is diagonal in
Fourier  space,  and  its  eigenvalues  $\lambda({\bf  k})$  have  the
limiting small-$k$ form
\begin{equation}
\lambda({\bf k}) \sim |{\bf k}|^{1+\mu},\ \ \ \ |{\bf k}| \to 0\ .
\label{spectrum}
\end{equation} 
In equation  (\ref{burgers}) we  have retained only  the leading-order
nonlinearity, which is  associated with the shear. In  this limit, the
noise correlator has  the same form as in  the zero-shear case, namely
(in  Fourier space) $\langle  \eta({\bf k},t)\eta({\bf  k},t') \rangle
\sim |{\bf  k}|^{\mu - 1}\delta({\bf  k}+{\bf k}')\delta(t-t')$, where
this particular form follows, via the fluctuation-dissipation theorem,
from   the  zero-shear  stationary   state,  $P[h({\bf   x})]  \propto
\exp[-{\rm  const}\, \int  d^dx (\nabla  h)^2]$.  The  parameter $\mu$
specifies      the      particular      dynamical     model      under
consideration. Particular  cases of physical relevance  are $\mu=1$ (a
nonconserved order  parameter, or `model  A' in the  classification of
Halperin and  Hohenberg \cite{Hohenberg}), $\mu=2$  (a conserved order
parameter  obeying  the Cahn-Hilliard  equation,  or  `model B'),  and
$\mu=0$ (a  conserved order parameter coupled to  hydrodynamic flow in
the viscous regime, or `model H').

The derivation  and RG analysis of equation  (\ref{burgers}) will form
the main  part of this work.   Since the system is  anisotropic due to
the shear,  we write ${\bf  x} = (x,{\bf  x}_\bot)$, where $x$  is the
coordinate  along  the  flow   direction,  and  ${\bf  x}_\bot$  is  a
$(d-2)$-dimensional vector  perpendicular to  the flow. There  are, in
general, three scaling exponents,  $\chi$, $\zeta$ and $z$, defined by
the condition that the  simultaneous scale transformations $x \to bx$,
${\bf x}_\bot \to b^\zeta {\bf x}_\bot$,  $h \to b^\chi h$, and $t \to
b^z t$ leave the interfacial  dynamics scale invariant. All three will
be determined  exactly for  all physical values  of $\mu$ and  for all
$d$.

The remainder of the paper  will consist of a more detailed discussion
of the  physical motivation for  these calculations, and  the analysis
and interpretation of the results. The interface equations are derived
in section  II for models  A, B, and  H.  Section III contains  the RG
analysis,  while in  section IV  we  discuss the  implications of  our
results for coarsening systems under shear. Section V concludes with a
summary of our results.

\section{The Interface Equation}

In each case we will  start from the relevant Ginzburg-Landau equation
for the  order parameter $\phi({\bf  r},t)$, and derive  the interface
equation by projecting the full equation of motion onto the interface.
We   assume   a   coarse-grained   free-energy   functional   of   the
Ginzburg-Landau form,
\begin{equation}
F[\phi] = \int  d^dr \left[\frac{1}{2}(\nabla\phi)^2 + V(\phi)\right]\
,
\end{equation}
where $V(\phi)$  is a symmetric  double-well potential with  minima at
$\phi=\pm 1$, representing the two equilibrium phases.

For  pedagogical purposes  we  begin  with the  simplest  case of  the
time-dependent Ginzburg-Landau equation (or `model A') which describes
phase-ordering in a system with a nonconserved scalar order parameter,
i.e.\ Ising-like systems such as a twisted-nematic liquid crystal.

\subsection{Model A}

We will consider  a uniform shear flow in  the $x$-direction, with the
velocity gradient in the  $y$-direction, ${\bf v}=\gamma y {\bf e_x}$,
where $\gamma$ is  the shear strength. The dynamics  of the the system
are governed by the Langevin equation
\begin{eqnarray}
\frac{\partial   \phi}{\partial   t}   +   \gamma   y   \frac{\partial
\phi}{\partial  x} &  = &  - \frac{\delta  F}{\delta \phi}  + \xi({\bf
r},t) \nonumber \\ &=& \nabla^2\phi - V'(\phi) + \xi({\bf r},t) \ ,
\label{modelA}
\end{eqnarray} 
where  the   second  term  on   the  left-hand  side  is   just  ${\bf
v}\cdot\nabla\phi$,  and   represents  the  advection   of  the  order
parameter by  the shear flow.   In equation (\ref{modelA}),  a kinetic
coefficient  has been  absorbed into  the timescale,  $V'(\phi) \equiv
dV/d\phi$, and $\xi({\bf r},t)$ is Gaussian white noise with mean zero
and correlator
\begin{equation}
\langle  \xi({\bf  r},t)\xi({\bf  r}',t')  \rangle  =  2D  \delta({\bf
r}-{\bf r}')\delta(t-t'),
\label{noise}
\end{equation}
where the noise strength $D$ is proportional to the temperature.

We now  construct an equation for  an interface, parallel  to the flow
direction  and  normal  to   the  velocity  gradient,  separating  the
equilibrium phases. We are interested in the limit where the interface
is almost planar,  such that $(\nabla h)^2$ is  typically small, i.e.\
we  are going  to systematically  neglect terms  which are  smaller by
powers of $(\nabla h)^2$ than the  terms we retain. In this limit, the
order parameter profile is well represented by the simple form
\begin{equation}
\phi({\bf r},t) = f[y - h({\bf x},t)]\ ,
\label{step}
\end{equation}
where we have written ${\bf r}  = ({\bf x},y)$. The function $f(u)$ is
essentially a  step function,  with a width  given by  the interfacial
width, $\xi_0$. Its derivative,  $f'(u)$, is therefore a smeared delta
function, which peaks on the  interface and has width $\xi_0$. It will
be used below as a projector onto the interface.

Substituting equation (\ref{step}) into equation (\ref{modelA}) gives,
with $u=y-h$,
\begin{eqnarray}
(\partial_t  h)f'(u)   &  =  &  -   \gamma(u+h)(\partial_x  h)f'(u)  -
[1+(\nabla h)^2]f''(u)  \nonumber \\ &  & + (\nabla^2 h)  f'(u)- V'(f)
\nonumber \\ & & + \xi[{\bf x}, u + h({\bf x},t),t]\ .
\label{stepA}
\end{eqnarray}
Finally   we  multiply   through   by  $f'(u)$   and  integrate   over
$u$. Formally we take the  integral from $-\infty$ to $\infty$, but in
practice  the   integral  is  concentrated  in   the  neighborhood  of
$u=0$. Since  $f''(u)f'(u)$ and $V'(f)f'(u)$  are perfect derivatives,
these terms drop out. Also the term involving $u[f'(u)]^2$ vanishes by
symmetry under the integral. The final result, therefore, is
\begin{equation}
\partial_t h + \gamma h \partial_x h = \nabla^2 h + \eta({\bf x},t).
\label{modelA-int}
\end{equation}
The noise term is given by
\begin{equation}
\eta({\bf   x},t)  =  -(1/\sigma)\int   du\,f'(u)\xi[{\bf  x},u+h({\bf
x},t),t]\ ,
\end{equation}
where $\sigma  = \int du\,[f'(u)]^2$  is the surface  tension. Clearly
the mean of $\eta$ is  zero, while use of equation (\ref{noise}) gives
its correlator as
\begin{equation}
\langle\eta({\bf x},t)\eta({\bf x'},t')\rangle=(2D/\sigma) \delta({\bf
x}-{\bf x}')\delta(t-t')\ .
\label{noiseA-int}
\end{equation}

In  the  zero-shear  limit,  $\gamma=0$,  equation  (\ref{modelA-int})
reduces  to the Edwards-Wilkinson  model \cite{ew},  and has  a simple
interpretation.   The interfacial  free energy  functional,  to lowest
order  in $(\nabla  h)^2$, is  $F_{int} =  (1/2)\int d^{d-1}x\,(\nabla
h)^2$.   The dynamics (\ref{modelA-int})  corresponds to  the Langevin
equation $\partial_t h  = -\delta F_{int}/\delta h +  \eta$. The noise
strength  $2D/\sigma$  in  (\ref{noiseA-int}) guarantees  the  correct
stationary distribution, $P[h] \propto \exp(-\sigma F[h]/D)$.

Before moving on to  model B, it is worth noting that  for the case of
zero   shear  and   zero  noise   the  equation   reduces   to  simple
relaxation. In Fourier space, one  has $\partial \tilde h({\bf k},t) =
-k^2 \tilde  h({\bf k},t)$,  i.e.\ fluctuations on  a length  scale $L
\sim 1/k$ relax  on a timescale $\tau(L) \sim  L^2$.  For a coarsening
system containing many interfaces,  this relation gives the timescale,
$L^2$, for  a feature  at scale  $L$ to relax  away, and  suggests the
relation  $L(t) \sim  t^{1/2}$  for the  coarsening  length scale,  or
`domain  scale'   in  a  phase-ordering  system.    This  approach  to
determining coarsening exponents from interfacial relaxation rates has
been used  before \cite{LTJZSO,shino}, and the  predictions agree with
the results  obtained from  other methods \cite{Review}.   Indeed, the
result is  more general \cite{Bray98}. In any  system where coarsening
proceeds by  relaxation of  extended defect structures  (domain walls,
vortex lines, etc.)  the dynamical exponent $z$, in the relation $L(t)
\sim  t^{1/z}$  for the  coarsening  dynamics,  is  the same  as  that
obtained  from  the  relaxation  rate, $\lambda({\bf  k})  \sim  |{\bf
k}|^z$, of a single defect  with a sinusoidal modulation at wavevector
${\bf k}$.  The  same general structure will be  apparent in the study
of models B and H.

\subsection{Model B}

For con\-ser\-ved dy\-na\-mics, the time-de\-pen\-dent Ginzburg-Landau
equation  is replaced  by the  Cahn-Hilliard-Cook equation  (i.e.\ the
noisy  Cahn-Hilliard equation)  which, in  the presence  of  a uniform
shear flow, reads
\begin{eqnarray}
\frac{\partial   \phi}{\partial   t}   +   \gamma   y   \frac{\partial
\phi}{\partial  x}  & =  &  \nabla^2  \frac{\delta  F}{\delta \phi}  +
\xi({\bf r},t)  \nonumber \\ &=& -\nabla^2[\nabla^2\phi  - V'(\phi)] +
\xi({\bf r},t) \ ,
\label{modelB}
\end{eqnarray} 
where   a   transport  coefficient   has   been   absorbed  into   the
timescale. The noise correlator is
\begin{equation}
\langle  \xi({\bf  r},t)\xi({\bf  r}',t')  \rangle  =  -  2D\,\nabla^2
\delta({\bf r}-{\bf r}')\delta(t-t')\ .
\label{noiseB}
\end{equation}
As a prelude to further analysis  it is convenient to first operate on
both sides of the equation  with the inverse of the Laplacian operator
(whose   meaning  will   become   clear  below).    Making  the   same
long-wavelength  approximation  (\ref{step}) as  in  the treatment  of
model A gives
\begin{eqnarray}
(-\nabla^2)^{-1}(\partial_t h + \gamma (u+h)\partial_x h) f'(u)
\hspace*{2.5cm} && \nonumber \\ = - [1+(\nabla h)^2]f''(u) + (\nabla^2
h) f'(u)-  V'(f) &&  \nonumber \\ +  (-\nabla^2)^{-1}\xi[{\bf x},  u +
h({\bf x},t),t]\ .\hspace*{1.7cm} &&
\end{eqnarray}
Multiplying through  by $f'(u)$, and  integrating over $u$  as before,
gives
\begin{eqnarray}
\int      du       f'(u)(-\nabla^2)^{-1}f'(u)[\partial_t      h      +
\gamma(u+h)\partial_x  h]  &  &  \nonumber  \\ =  \sigma\nabla^2  h  +
\bar{\eta}({\bf x},t)\ , &&
\label{B1}
\end{eqnarray}
where the noise is given by
\begin{equation}
\bar{\eta}({\bf  x},t)  =  - \int  du\,f'(u)(-\nabla^2)^{-1}  \xi[{\bf
x},u+h({\bf x},t),t]\ .
\label{noiseB-int}
\end{equation}

The  meaning of  the operator  $(-\nabla^2)^{-1}$ is  as  follows.  In
Fourier space  one has $(-\nabla^2)^{-1} \to (k^2  + q^2)^{-1}$, where
$({\bf k},q)$ is the vector conjugate to $({\bf x},y)$.  Defining, for
a  general  function  $F$,  $G({\bf x},y)  =  (-\nabla^2)^{-1}  F({\bf
x},y)$,  its Fourier  transform, in  the  $(d-1)$-dimensional subspace
spanned by ${\bf x}$, is given by
\begin{equation}
\tilde{G}({\bf k},y) = \frac{1}{2|{\bf k}|}\int_{-\infty}^\infty dy'\,
\exp(-|{\bf k}|\,|y-y'|)\,\tilde{F}({\bf k},y')\ .
\end{equation}
We  now  use  this  result  to  evaluate the  left  side  of  equation
(\ref{B1}). The leading  order non-linearity (in $h$) is  given by the
shear  term,  so  elsewhere  in  equation (\ref{B1})  we  neglect  the
distinction  between  $u$ and  $y=u+h$.   It  can  be shown  that  the
leading-order   terms  omitted   in   this  approach   are  of   order
$h(\partial_xh)^2$. Denoting, for  brevity, the Fourier transform with
respect to ${\bf  x}$ by a subscript ${\bf  k}$, the Fourier transform
of the left-side of (\ref{B1}) becomes
\begin{eqnarray}
\frac{1}{2|{\bf k}|} \int du\int dv\,\exp(-|{\bf k}|\,|u-v|)f'(u)f'(v)
&&  \nonumber  \\ \times  \left[\partial_t  h_{\bf  k}  + i\gamma  k_x
\left(v+\frac{1}{2}[h^2]_{\bf k}\right)\right]. &&
\end{eqnarray} 
Recalling  that  $f'(u)$ acts  like  a  delta  function at  $u=0$  (of
strength 2, which  is the discontinuity of the  order parameter across
the interface) equation (\ref{B1}) simplifies to
\begin{equation} 
\partial_t  h_{\bf k}  +  \frac{i}{2}  \gamma k_x  [h^2]_{\bf  k} =  -
\frac{\sigma}{2}|{\bf    k}|^3\,h_{\bf     k}    +    \frac{1}{2}|{\bf
k}|\,\bar{\eta}_{\bf k}(t)\ .
\label{B2}
\end{equation}

Consider  once more  the  case of  zero  shear and  zero noise.   Then
equation (\ref{B2}) represents simple relaxation, with fluctuations on
length scale  $L \sim 1/k$  relaxing at a  rate $k^3$, i.e.\  as $k^z$
with $z=3$.  This is again consistent with the known coarsening growth
law, $L(t) \sim t^{1/3}$, for model B \cite{Review}.

The  form of  the  noise  correlator can  be  extracted from  equation
(\ref{noiseB-int}). Using  the same simplifications  as before yields,
in Fourier space,
\begin{equation}
\langle   \bar{\eta}_{\bf  k}(t)\bar{\eta}_{-{\bf   k}'}(t')\rangle  =
 \frac{4D}{|{\bf k}|}\,\delta_{{\bf k},{\bf k}'}\,\delta(t-t')\ .
\label{noiseB1-int}
\end{equation}

Equation (\ref{B2}) has, in real space, precisely the form of equation
(\ref{burgers}),  where the  operator $\cal{L}$  has  the small-$|{\bf
k}|$ spectrum  $\lambda({\bf k}) \sim  |{\bf k}|^3$, i.e.\ it  has the
form  (\ref{spectrum})  with  $\mu=2$.   Defining $\eta_{\bf  k}(t)  =
\frac{1}{2}  |{\bf k}|\bar{\eta}_{\bf  k}(t)$,  one recovers  equation
(\ref{burgers}) exactly, with noise correlator
\begin{equation}
\langle\eta_{\bf    k}(t)\eta_{-{\bf    k}'}(t')\rangle    =    D|{\bf
 k}|\,\delta_{{\bf k},{\bf k}'}\,\delta(t-t')\ .
\label{noiseB2-int}
\end{equation}

For   model  A,   equation  (\ref{modelA-int})   also  has   the  form
(\ref{burgers}), but  with $\mu=1$ in  (\ref{spectrum}). This suggests
that both models be viewed as members of a more general class, defined
by   equations  (\ref{burgers})   and   (\ref{spectrum})  with   $\mu$
general. As  discussed in the  Introduction, the requirement  that the
equilibrium distribution $P[h] \propto \exp[-\frac{\sigma}{2}\sum_{\bf
k} k^2 h_{\bf k} h_{-{\bf k}}]$ be recovered for $\gamma=0$ forces the
noise correlator to have  the form $ \langle\eta_{\bf k}(t)\eta_{-{\bf
k}'}(t')\rangle    \sim   |{\bf    k}|^{\mu-1}\,\delta_{{\bf   k},{\bf
k}'}\,\delta(t-t')$.      Our    results     (\ref{noiseA-int})    and
(\ref{noiseB2-int}),  for models  A and  B respectively,  satisfy this
requirement.

\subsection{Model H} 

The general results relating the form of the spectrum (\ref{spectrum})
of the operator  $\cal{L}$ in (\ref{burgers}) to the  exponent $z$ for
coarsening ($L(t)  \sim t^{1/z}$),  and the form  of the noise  to the
requirement of recovering the correct equilibrium state in zero shear,
suggests a simple form for the  equation of motion for an interface in
a phase-separating binary fluid  in the `viscous hydrodynamic' regime.
This is  the regime described  by `model H' of  the Hohenberg-Halperin
scheme \cite{Hohenberg}.  In this  regime, it is known that coarsening
proceeds  linearly in  time,  $L(t) \sim  t$,  corresponding to  $z=1$
\cite{Siggia}. This suggests  that the interfacial relaxation spectrum
is given  by $\lambda({\bf k})  \sim |{\bf k}|$  for ${\bf k}  \to 0$,
i.e.\ $\mu=0$  in (\ref{spectrum}), a result which  has been confirmed
by Shinozaki \cite{shino}.  This in turn suggests that the interfacial
noise correlator should have the small ${\bf k}$ form corresponding to
$\mu=0$, namely $\langle \eta_{\bf k}(t) \eta_{-{\bf k}'}(t')\rangle =
D|{\bf k}|^{-1}\delta_{{\bf k},{\bf k}'} \delta(t-t')$.

We now show that  these expectations, based on general considerations,
are indeed borne out in practice. In the absence of thermal noise, the
equation of motion for the order parameter field takes the form
\begin{equation}
\frac{\partial\phi}{\partial  t}  +  {\bf v}\cdot\nabla\phi  =  \Gamma
 \nabla^2\mu\
\label{modelH}
\end{equation} 
where  $\mu =  \delta F/\delta  \phi$  is the  chemical potential  and
$\Gamma$ is a  transport coefficient. The velocity, ${\bf  v}$, of the
fluid, assumed incompressible, satisfies the Navier-Stokes equation
\begin{equation}
\rho\left(\frac{\partial{\bf      v}}{\partial     t}      +     ({\bf
v}\cdot\nabla){\bf  v}\right) =  \eta \nabla^2  {\bf v}  - \nabla  p -
\phi\nabla\mu\ ,
\label{NS}
\end{equation}
where $\rho$  and $\eta$  are the density  and viscosity of  the fluid
respectively, and  $p$ is the  pressure. The final term  in (\ref{NS})
contains  the  feedback between  the  order  parameter  and the  fluid
velocity.

The  coarsening dynamics  of this  system  is known  to exhibit  three
regimes \cite{Siggia,Furukawa}: (i)  an early time `diffusive' regime,
where  the hydrodynamics  is irrelevant  (the fluid  velocity  is much
smaller than the typical interface  velocity) and the model reverts to
model  B,   with  coarsening  scale  $L(t)  \sim   t^{1/3}$;  (ii)  an
intermediate time  `viscous hydrodynamic' regime,  where the `inertial
terms' on the left side  of equation (\ref{NS}) can be neglected, with
$L(t) \sim t$; (iii) a  late time `inertial hydrodynamic' regime where
the inertial  terms dominate the  viscous term, $\eta  \nabla^2v$, and
$L(t) \sim t^{2/3}$.
 
Here we focus on the viscous hydrodynamic regime, where we can set the
left   side   of  (\ref{NS})   to   zero.    This   defines  model   H
\cite{Hohenberg,Review}.  For  simplicity, we will  ignore the imposed
shear flow in  the first instance.  The pressure  can be eliminated by
using the  incompressibility condition,  $\nabla \cdot {\bf  v}=0$, to
express the velocity in  terms of $\phi\nabla\mu$.  Putting the result
into (\ref{modelH}), and adding a  noise term gives the final equation
for model H. Since we are  interested in the regime where diffusion is
negligible, we drop the term $\Gamma\nabla^2\mu$ to obtain
\begin{equation}
\frac{\partial\phi}{\partial    t}    =    -    \int    {\bf    dr}'\,
\partial_a\phi({\bf     r})\,T_{ab}({\bf    r}    -     {\bf    r}')\,
\partial_b\phi({\bf r}')\,\mu({\bf r}') + \xi({\bf r},t)\ ,
\label{modelH2}
\end{equation} 
where  $\mu  =   \delta  F/\delta\phi  =  V'(\phi)-\nabla^2\phi$,  and
$T_{ab}$ is the Oseen tensor, with Fourier transform
\begin{equation}
T_{ab}({\bf   k})   =   \frac{1}{\eta   k^2}\,   \left(\delta_{ab}   -
\frac{k_ak_b}{k^2}\right)\ .
\label{Oseen}
\end{equation}
In equation  (\ref{modelH2}), repeated  indices are summed  over.  The
form    of    the    noise    correlator   is    dictated    by    the
fluctuation-dissipation theorem:
\begin{eqnarray}
\langle\xi({\bf    r},t)\xi({\bf   r}',t')\rangle    &   =    &   2D\,
\partial_a\phi({\bf     r})\,T_{ab}({\bf    r}    -     {\bf    r}')\,
\partial_b\phi({\bf   r}')   \nonumber   \\   &  &   \times   \mu({\bf
r}')\delta(t-t')\ ,
\label{noiseH} 
\end{eqnarray}
where $D$ is the temperature.

To determine  the interface equation  we insert the  form (\ref{step})
into (\ref{modelH2}) to obtain, analogous to (\ref{stepA})
\begin{eqnarray} 
(\partial_t  h)f'(u)  &  =  &  \int  {\bf  dr}'\,  \partial_a\phi({\bf
r})\,T_{ab}({\bf r}  - {\bf r}')\,  \partial_b\phi({\bf r}') \nonumber
\\ && \times \left\{(\nabla^2h)f'(v) + V'[f(v)]\right. \nonumber \\ &&
\left.- [1+(\nabla h)^2]f''(v)\right\} \nonumber \\ && - \xi({\bf x},u
+ h({\bf x},t),t)\ ,
\label{stepH}
\end{eqnarray} 
where $u=y-h({\bf x},t)$ and  $v=y'-h({\bf x'},t)$. It is important to
note that  the Oseen tensor in  real space is only  defined for $d>2$.
Therefore, all the following equations  for model H are only valid for
$d>2$.

As in models  A and B, the  leading term for small $h$  comes from the
$\nabla^2 h$ term  in the braces.  To linear  order, therefore, we can
use a `flat interface approximation'  in the terms outside the braces.
This  means we  can write  $\nabla  \phi({\bf r)}  = f'(u){\bf  e}_y$,
$\nabla\phi({\bf r}') =  f'(v){\bf e}_y$, where ${\bf e}_y$  is a unit
vector in  the $y$ direction,  and $T_{yy}$ becomes the  only relevant
element of  the Oseen tensor. Multiplying both  sides of (\ref{stepH})
by $f'(u)$, and integrating over $u$, yields, to leading order in $h$,
\begin{equation}
\partial_t  h({\bf x})  = \sigma  \int {\bf  dx}'\,T_{yy}({\bf x}-{\bf
x}',0)\, \nabla^2 h({\bf x}') + {\rm noise}\ ,
\end{equation}
where the integral  is over the $(d-1)$-dimensional plane  of the mean
interface.   Fourier transforming  this  result, using  (\ref{Oseen}),
gives
\begin{equation}
\frac{\partial   h_{\bf   k}}{\partial   t}  =   -\frac{\sigma   |{\bf
k}|}{4\eta}\,h_{\bf k} + \eta_{\bf k}(t)\ ,
\label{modelH-int}
\end{equation}
where ${\bf  k}$ is  now a $(d-1)$-dimensional  vector, and  we recall
that $d>2$.  The  noise correlator can by evaluated  by exploiting the
`flat interface'  limit, valid to  leading (zeroth) order in  $h$. The
result is
\begin{equation}
\langle      \eta_{\bf     k}(t)\eta_{-{\bf      k}'}(t')\rangle     =
 \frac{D}{2\eta|{\bf k}|}\,\delta_{{\bf k},{\bf k}'}\,\delta(t-t')\ .
\label{noiseH-int}
\end{equation}
Equations (\ref{modelH-int}) and (\ref{noiseH-int}) have precisely the
forms anticipated  earlier on  general grounds. We  note that,  in the
absence  of thermal noise,  our approach  is very  similar to  that of
Shinozaki \cite{shino}.

Finally, we have  to impose the shear flow. To do  this we write ${\bf
v} = \gamma y  {\bf e}_x + {\bf u}$, where ${\bf  u}$ is the deviation
from   the  mean   shear  flow   and  should   vanish  far   from  the
interface. Inserting  this form for  ${\bf v}$ in  both (\ref{modelH})
and  (\ref{NS}),  with  the  left  side  of  (\ref{NS})  set  to  zero
appropriate to the  viscous regime, we find that  the shear term drops
out  of  both the  Navier-Stokes  equation  and the  incompressibility
condition. We conclude  that ${\bf u}$ plays exactly  the same role in
the sheared  case as ${\bf v}$  plays in the unsheared  case, and that
the effect of the imposed shear  is to add a term $\gamma y \partial_x
\phi$ to the left side of  (\ref{modelH2}), just as in models A and B,
and therefore a term $(i/2)\gamma  k_x [h^2]_{\bf k}$ to the left side
of (\ref{modelH-int}), which then becomes
\begin{equation}
\frac{\partial  h_{\bf  k}}{\partial   t}  +  \frac{i}{2}  \gamma  k_x
[h^2]_{\bf k} = -\frac{\sigma |{\bf k}|}{4\eta}\,h_{\bf k} + \eta_{\bf
k}(t)\ .
\label{modelH-int2}
\end{equation}

\section{Renormalization Group Analysis}
The    starting   point    of    the   RG    analysis   is    equation
(\ref{burgers}). Since,  however, the system is  anisotropic we expect
difference   scaling  properties  in   the  directions   parallel  and
perpendicular to  the shear. Under coarse  graining, anisotropies will
develop in the  linear terms in the equation.   Additionally, from the
structure  of the  non-linear  (shear)  term it  is  clear that  terms
analytic  in  $k_x^2$  will  be  generated in  the  response  function
self-energy  and  the   renormalized  noise.   Anticipating  this,  we
generalize equation (\ref{burgers}) to (in Fourier space):
\begin{equation}
\partial_t h_{\bf k} + \frac{i}{2}\gamma k_x (h^2)_{\bf k} = -(\lambda
|{\bf k}|^{1+\mu} + \nu_xk_x^2) h_{\bf k} + \eta_{\bf k}(t)\ .
\label{burgers1}
\end{equation} 
The noise correlator takes the form
\begin{equation}
\langle\eta_{\bf    k}(t)\eta_{-{\bf    k}'}(t')\rangle   =    (D|{\bf
 k}|^{\mu-1} + D_xk_x^2)\, \delta_{{\bf k},{\bf k}'}\,\delta(t-t')\ .
\label{noiseRG-int}
\end{equation}

We apply a  momentum-shell RG in which, for  convenience, we impose an
ultraviolet   momentum  cut-off,   $\Lambda$,  in   the  $x$-direction
only. The  RG transformation consists of three  steps: (i) eliminating
modes with $\Lambda/b < |k_x|  < \Lambda$ (hard modes); (ii) rescaling
the length scales, $x$ and  $\bf x_\bot$, the field variable, $h$, and
the  time, $t$;  (iii) looking  for fixed  points of  the  equation of
motion at which the theory is  invariant under (i) and (ii). As usual,
the  elimination of  modes will  be executed  perturbatively  near the
critical dimension, $d_c$,  of the theory. We will  show that $d_c$ is
given by $d_c = (9+\mu)/2$ for  $\mu \ge 1$, while for $\mu<1$ we will
see that the situation is less clear.

The scale transformation takes the form
\begin{equation}
x=bx',\ \ {\bf x}_\bot = b^\zeta{\bf  x}_\bot' ,\ \ h= b^\chi h',\ \ t
= b^z t'\ .
\label{scale}
\end{equation}
To make further progress it is necessary to know whether $\zeta \le 1$
or $\zeta > 1$. Since the  shear term tends to enhance the interfacial
coarsening in the $x$-direction,  we expect to find $\zeta<1$ whenever
the shear  is relevant,  though $\zeta =  1$ is possible  for $d>d_c$,
where the shear  rate $\gamma$ is formally an  irrelevant variable. We
will further argue that  $\zeta>1$ is unphysical, and will accordingly
restrict consideration  to $\zeta  \le 1$ in  the following.   We will
find,  however, that  the nature  of  the theory  for $d>d_c$  differs
according  to whether  $\mu \ge  1$ or  $\mu <  1$. We  will therefore
consider  these two  regimes  separately. The  former regime  includes
models A ($\mu=1$) and B  ($\mu=2$), while the latter includes model H
($\mu=0$). A  brief discussion,  in the present  context, of  the case
$\mu=1$ can be found in  \cite{brief}. This special case had also been
discussed  earlier in  the (physically  very different)  context  of a
sandpile model \cite{hwa}.

\subsection{The case $\mu \ge 1$}

A  value  of $\zeta$  less  than  unity  implies anisotropic  scaling.
Furthermore,  in such cases  the transverse  part, ${\bf  k}_\bot$, of
${\bf k}$ dominates over $k_x$ in the terms involving powers of $|{\bf
k}|$, both in  the equation of motion and  the noise correlator, which
then take the following forms:
\begin{equation}
\partial_t h_{\bf k} + \frac{i}{2}\gamma k_x (h^2)_{\bf k} = -(\lambda
|{\bf k}_\bot|^{1+\mu} + \nu_xk_x^2) h_{\bf k} + \eta_{\bf k}(t)\ . \\
\label{burgers2}
\end{equation}
\begin{equation}
\langle\eta_{\bf    k}(t)\eta_{-{\bf    k}'}(t')\rangle   =    (D|{\bf
 k}_\bot|^{\mu-1}      +     D_xk_x^2)\,      \delta_{{\bf     k},{\bf
 k}'}\,\delta(t-t')\ .
\label{noiseRG-int2}
\end{equation}
Note that  for $\mu=1$ the  term $\lambda k_x^2$ coming  from $\lambda
|{\bf k}|^2$  can be  absorbed into the  $\nu_xk_x^2$ term,  while the
term $D|{\bf k}|^{\mu-1}$  becomes a constant. So the  case $\mu=1$ is
covered  by the  general structure  of equations  (\ref{burgers2}) and
(\ref{noiseRG-int2}).

Applying the transformation (\ref{scale}) to equation (\ref{burgers2})
then yields rescaled values for the parameters in the equation and the
noise correlator:
\begin{eqnarray}
\gamma' & = & b^{\chi+z-1}\,\gamma
\label{gamma} \\
\lambda' & = & b^{z-(1+\mu)\zeta}\,\lambda
\label{lambda} \\
\nu_x' & = & b^{z-2}\,\nu_x + \cdots
\label{nu} \\
D' & = & b^{z-2\chi-1 - (\mu-1)\zeta - (d-2)\zeta}\,D
\label{D} \\
D_x' & = & b^{z-2\chi-3 -(d-2)\zeta}\,D_x + \cdots ,
\label{Dx}
\end{eqnarray}
where  the  ellipses indicate  that  the  parameters  $\nu$ and  $D_x$
acquire perturbative  corrections due  to the coarse-graining  step of
the RG procedure. By  contrast, the parameters $\gamma$, $\lambda$ and
$D$   acquire   {\em  no}   perturbative   corrections  --   equations
(\ref{gamma}), (\ref{lambda}) and (\ref{D})  are exact. The absence of
perturbative corrections  to $\gamma$  follows from the  invariance of
the   general   equation  of   motion,   (\ref{burgers}),  under   the
transformation $h \to  h+h_0$, $x \to x + \gamma h_0  t$, which is the
analog  for our  system of  the usual  Galilean invariance  of Burgers
equations (see, for example,  \cite{fns}).  The absence of corrections
to (\ref{lambda}) and (\ref{D}) follows  from the fact that the vertex
$\gamma$  carries  a factor  $k_x$.   As  a  result, all  perturbative
contributions  to  the response  function  self-energy  and the  noise
correlator carry factors of $k_x^2$.

Let us  first examine the  linear theory ($\gamma=0$) to  identify the
critical  dimension  $d_c$.   In  the  linear  theory,  there  are  no
perturbative corrections, and equations (\ref{lambda})--(\ref{Dx}) all
hold exactly. From (\ref{lambda})--(\ref{D}) we obtain
\begin{equation}
z_0   =  2\  ,\   \  \zeta_0   =  \frac{2}{1+\mu}\   ,\  \   \chi_0  =
\frac{7-\mu-2d}{2(1+\mu)}\ ,
\label{free1}
\end{equation} 
where the subscripts  indicate that these are the  results of the free
theory. Inserting these exponents into equation (\ref{Dx}) gives $D_x'
=  b^{-4/(1+\mu)}D_x$, indicating  that $D_x$  flows to  zero  at this
fixed point.

Equation (\ref{gamma}) determines the  relevance, at the trivial fixed
point,  of the  shear  rate $\gamma$.   From  (\ref{free1}) we  obtain
$\chi_0 + z_0 - 1 = (9+\mu-2d)/[2(1+\mu)]$. Hence $\gamma$ is relevant
for $d<d_c$, where
\begin{equation}
d_c = (9+\mu)/2\ ,\ \ \ \mu \ge 1\ .
\label{dc1}
\end{equation}
For  $d <  d_c$,  we expect  a new  fixed  point to  appears at  which
$\gamma$, $\lambda$ and $D$ are all non-zero. Equations (\ref{gamma}),
(\ref{lambda}) and (\ref{D}) give the corresponding exponents exactly:
\begin{equation}
z =  \frac{3(1+\mu)}{6+2\mu-d}\ ,\ \ \zeta =  \frac{3}{6+2\mu-d}\ ,\ \
\chi = \frac{3-\mu-d}{6+2\mu-d}\ .
\label{nontrivial}
\end{equation}
We recall that  in order for our calculation to  be consistent we must
have  $\zeta\leq 1$,  such that  $|\bf k|  \sim |\bf  k_\perp|$.  From
relations  (\ref{free1})  and  (\ref{nontrivial})  we  see  that  this
condition  requires $\mu  \geq 1$,  consistent  with the  case we  are
currently analyzing.

Exponents  (\ref{nontrivial})  are correct  only  if  the fixed  point
values of the parameters $\gamma$, $\lambda$ and $D$ are all non-zero,
otherwise their  scaling dimensions  cannot be set  equal to  zero. To
check this  fact we perform a  one-loop RG calculation  to compute the
perturbative corrections to $\nu_x$.  In general, integration over the
hard  modes   gives  the  following  equation   for  the  renormalized
propagator $G^<({\bf k},\omega)$ (see Fig.1): \beq G^<({\bf k},\omega)
=     G({\bf    k},\omega)     +     G({\bf    k},\omega)\,\Sigma({\bf
k},\omega)\,G^<({\bf k},\omega) \ , \eeq
\begin{figure}
\begin{center}
\leavevmode \epsfxsize=3in \epsffile{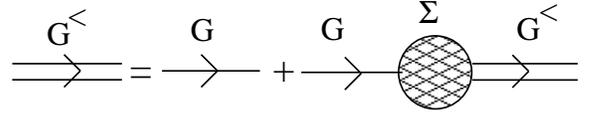}
\end{center}
\caption{Dyson  equation  for the  propagator  in  terms  of the  bare
propagator (single lines) and the self-energy (hatched circle).}
\end{figure}
\noindent
where the  bare propagator is  given by 
\beq  G({\bf k},\omega)^{-1}=
-i\omega + \nu_x  k_x^2 + \lambda |{\bf k}_\perp|^{1+\mu} \  , 
\eeq 
and the self-energy $\Sigma({\bf   k},\omega)$  must   be   calculated
perturbatively  in   $\gamma$.   From  the   relation,  \beq  G^<({\bf
k},\omega)^{-1} = G({\bf k},\omega)^{-1}-\Sigma({\bf k},\omega) \ ,
\label{bau}
\eeq we clearly see that  the perturbative corrections to $\nu_x$ come
from  terms of  order  $k_x^2$ in  $\Sigma({\bf k},\omega)$.   Setting
$b=e^l$, with $l$  infinitesimal, equations (\ref{nu}) and (\ref{bau})
yield,  \beq \frac{d\nu_x}{dl}= \nu_x\left[  (z-2) -  \lim_{{\bf k}\to
0}\, \frac{1}{\nu_x \, k_x^2\, l}\, \Sigma({\bf k},0)\right] \ .
\label{nunu}
\eeq The  standard one-loop  diagram for the  self-energy is  shown in
Figure 2 (see, for example, \cite{fns}).
\begin{figure}
\begin{center}
\leavevmode \epsfxsize=3in \epsffile{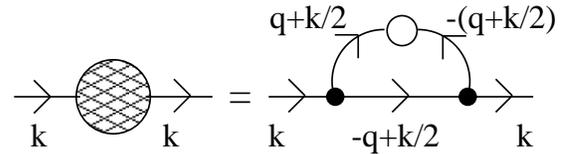}
\end{center}
\caption{One-loop contribution to  the self-energy. The internal lines
also carry frequency labels (not shown).}
\end{figure}
\noindent
Full  circles  represent  $\gamma$-vertices,  open  circles  represent
contractions  of  the  noise,  $\langle  \eta_{\bf  k}  \eta_{-\bf  k}
\rangle\sim D$, and  arrows are bare propagators. The  leading term of
the self-energy in the limit $({\bf k}, \omega)\to 0$ is given by
\end{multicols}
\beq   \Sigma({\bf   k},0)    =   -\gamma^2   D   \int_{\Omega,   q^>}
k_x\left(\frac{k_x}{2}-q_x\right)  \left|G\left(\frac{{\bf k}}{2}+{\bf
q},\Omega\right)\right|^2         G\left(\frac{{\bf        k}}{2}-{\bf
q},\Omega\right) \left|\frac{{\bf k}_\perp}{2}+{\bf q}_\perp\right| =
\label{sissy}\\ - \frac{2(6+2\mu-d)}{\mu+5-d} \, U \ \nu_x \, k_x^2\,l    
\eeq   
\beq   U=\frac{S_{d-2}}{8(\mu+1)(2\pi)^{d-2}}   \,
\Gamma\left(\frac{d+\mu-3}{\mu+1}\right)
\Gamma\left(\frac{6+2\mu-d}{\mu+1}\right)    \,    \gamma^2    D    \,
\lambda^\frac{3-d-\mu}{\mu+1} \, \nu_x^{-\frac{6+2\mu-d}{\mu+1}} \ .
\label{effective}
\eeq
\begin{multicols}{2}
\noindent

In  the  expression  above  $\Gamma(u)$  is  the  gamma  function  and
$S_{d-2}$ is the surface area  of the unit sphere in $d-2$ dimensions.
The notation $(\Omega, q^>)$ means  that we integrate with the measure
$d\Omega\, dq_x\, d{\bf q}_\perp/(2\pi)^{d-1}$, within the outer shell
$\Lambda e^{-l} <  |q_x| <\Lambda$.  Due to the  anisotropic nature of
the non-linearity, there  is no need to introduce  a cut-off for ${\bf
q}_\perp$.   Furthermore, we  have taken  $\Lambda=1$ without  loss of
generality.

Putting    together   equations   (\ref{nunu}),    (\ref{sissy})   and
(\ref{effective}), and using the  scaling dimensions of the parameters
$\gamma, \lambda$ and $D$, we  finally obtain the RG flow equation for
the   effective   coupling   constant   $U$,  \beq   \frac{dU}{dl}   =
\frac{9+\mu-2d}{\mu+1}\,U -\frac{2(6+2\mu-d)^2}{(\mu+1)(\mu+5-d)}\,U^2
\ .
\label{superflow}
\eeq  Consistent  with  our  previous determination  of  the  critical
dimension, we  see that the linear  term in the  flow equation changes
sign for $d=d_c=(9+\mu)/2$. Moreover,  the quadratic term is negative,
implying that for  any $d<d_c$ there is a  non-zero stable fixed point
$U^\star=O(\epsilon)$,  with  $\epsilon=d_c-d$.   The RG  perturbative
expansion is thus well behaved  and the fixed point values of $\gamma,
\lambda$   and   $D$   for   $d<d_c$  are   finite.    The   exponents
(\ref{nontrivial})  are therefore  correct.   On the  other hand,  for
$d>d_c$ the  only stable fixed point is  $U^\star=0$, corresponding to
an irrelevant  non-linearity and thus  giving the `free'  exponents of
(\ref{free1}).

\subsection{The case $\mu < 1$}

For $\mu <  1$, equation (\ref{free1}) gives $\zeta >  1$ for the free
theory, violating the assumption $\zeta < 1$ under which (\ref{free1})
was derived. This suggests we look  for a solution with $\zeta \ge 1$.
In this case, $k_x$ will dominate over (or be the same order as) ${\bf
k}_\bot$  in  $|{\bf k}|$.   The  recursion  relations for  $\lambda$,
$\nu_x$ and $D$ become
\begin{eqnarray}
\lambda' & = & b^{z-(1+\mu)}\,\lambda
\label{lambda1} \\
\nu_x' & = & b^{z-2}\,\nu_x + \cdots
\label{nu1} \\
D' & = & b^{z-2\chi-1 - (\mu-1) - (d-2)\zeta}\,D
\label{D1}
\end{eqnarray}
instead of  (\ref{lambda})--(\ref{D}). At the fixed point  of the free
theory ($\gamma=0$), equation (\ref{lambda1}) gives $z=1+\mu$, so that
(\ref{nu1})  becomes  $\nu_x' =  b^{\mu  -1}\nu_x$,  i.e.\ $\nu_x$  is
driven  to zero, since  $\mu<1$. The  theory with  $\nu_x=0=\gamma$ is
completely isotropic, so  $\zeta=1$. Inserting $z=1+\mu$ and $\zeta=1$
in (\ref{D1}) gives $\chi=(3-d)/2$.  Summarising, the exponents of the
free theory for $\mu < 1$ are
\begin{equation}
z_0 = 1+\mu\ ,\ \ \zeta_0 = 1\ ,\ \ \chi_0 = (3-d)/2\ ,
\label{free2}
\end{equation}
which coincides with (\ref{free1}) in the limit $\mu \to 1$.

The   relevance  of   $\gamma$   is  again   determined  by   equation
(\ref{gamma}).   From  (\ref{free2}),  the combination  $\chi+z-1$  is
given, for  the free theory, by $\chi_0+z_0-1  = (3+2\mu-d)/2$. Hence,
for $\mu<1$, $\gamma$ is relevant below the new critical dimension
\begin{equation}
d_c' = 3 + 2\mu\ ,\ \ \ \mu<1\ .
\label{dc2}
\end{equation}
Note that $d_c'$  differs from the critical dimension  $d_c$ found for
the case $\mu \geq 1$ in (\ref{dc1}), namely $d_c'<d_c$.  On the other
hand they coincide in the limit $\mu \to 1$.

For $d<d_c'$, from (\ref{nontrivial})  one again obtains $\zeta<1$ and
therefore  it is  tempting  to  conclude that  these  are the  correct
exponents even  for the  $\mu<1$ case, provided  that $d<d_c'$.   As a
further consistency check, one may note that these exponents reproduce
the ones of  the free theory given by (\ref{free2})  for $d \to d_c'$.
Unfortunately, the situation is not as simple as this. If we perform a
one-loop perturbative expansion below $d_c'$, we formally get the same
flow   equation    (\ref{superflow}),   since   $\zeta<1$    in   this
regime. However, as we have seen,  the fixed point of this equation is
of  order  $\epsilon=d_c-d$,  which  is  {\it not}  small  for  $d\sim
d_c'$. In  other words, because of  the gap between  $d_c'$ and $d_c$,
the one-loop expansion  in the form stated above  is not under control
in  the regime $\mu<1$.   We were  not able  to find  a perturbatively
consistent  solution in  this phase.   As a  consequence, we  can only
conjecture that the  exponents we have found for  $\mu<1$ are correct,
since they lack a substantial perturbative support.

Finally, let  us note that although  one can formally  find a solution
with $\zeta>1$,  all the  terms involving ${\bf  k}_\bot$ drop  out at
this   fixed    point,   and   the    equation   becomes   essentially
one-dimensional,  which  is  unphysical.   We  therefore  reject  this
possibility.

\subsection{The case $d=2$}

Some of the results derived above only hold for $d>2$. This is because
the idea that ${\bf k}_\bot$ dominates $k_x$ in $|{\bf k}|$ is clearly
inapplicable  in $d=2$,  since  there is  only  $k_x$. Similarly,  the
exponent  $\zeta$ can  no longer  be defined,  so there  are  just two
independent  exponents, $z$ and  $\chi$.  The  equation of  motion and
noise    correlator     are    given    by     (\ref{burgers2})    and
(\ref{noiseRG-int2})  respectively, but with  $|{\bf k}|$  replaced by
$|k_x|$. We recall that model H ($\mu=0$) is ill-defined for $d=2$.

The RG recursion relations for $d=2$ become,
\begin{eqnarray}
\gamma' & = & b^{\chi+z-1}\,\gamma
\label{gamma2} \\
\lambda' & = & b^{z-\mu-1}\,\lambda
\label{lambda2} \\
\nu_x' & = & b^{z-2}\,\nu_x + \cdots
\label{nu2} \\
D' & = & b^{z-2\chi-\mu}\,D
\label{D2} \\
D_x' & = & b^{z-2\chi-3}\,D_x + \cdots ,
\label{Dx2}
\end{eqnarray}
Equations  (\ref{gamma2}), (\ref{lambda2})  and (\ref{D2})  are exact,
and  therefore it  seems that  we have  three equations  for  just two
unknown exponents, $\chi$ and $z$.  This apparent paradox is solved if
one of the  parameters is zero at the fixed point,  since in this case
the corresponding equation is  trivially satisfied without setting the
scaling  dimension  to zero,  The  shear  rate  $\gamma$ is  certainly
relevant, since  $d=2$ is below  the critical dimension. If  we assume
$D\to 0$,  using equations  (\ref{gamma2}) and (\ref{lambda2})  we get
$\chi+z=1$  and $z=\mu+1$,  giving  $\chi=-\mu$.  This  would imply  a
positive scaling  dimension for $D$, which is  inconsistent with $D\to
0$.   Thus, we  must assume  $\lambda\to 0$,  and find  from equations
(\ref{gamma2})  and (\ref{D2}),  $\chi+z=1$ and  $z-2\chi=\mu$  at the
fixed point, giving
\begin{equation}
z = (2+\mu)/3\ ,\ \ \ \chi = (1-\mu)/3
\end{equation}
in $d=2$. Inserting these results into (\ref{lambda2}) gives $\lambda'
=  b^{-(1+2\mu)/3}\lambda$, so $\lambda$  flows to  zero in  $d=2$, as
assumed, for all $\mu >-1/2$.

\section{stability of the domains}

The calculations of the previous  sections are important to assess the
stability of the highly stretched domains in a coarsening system under
shear.  We  recall that  we are considering  a shear  velocity profile
with flow in  the $x$ direction and gradient in  the $y$ direction. We
denote by $\bf x_\perp$ all  the directions orthogonal to both $x$ and
$y$  for  $d\geq 3$.   The  effect  of the  shear  is  to stretch  the
coarsening domains,  such that there are two  different length scales,
$L_\parallel$,  along the  $x$  direction, and  $L_\perp$  in all  the
orthogonal directions. The transverse  size of the domains, $L_\perp$,
is  in  general  much  smaller than  longitudinal  one,  $L_\parallel$
\cite{experiment,theory}.  What  we have to check is  whether the size
$\Delta$ of the height  fluctuation is larger than $L_\perp$, inducing
a breaking of the domains, or whether $\Delta < L_\perp$, meaning that
the domains are stable under thermal fluctuations.

In the  long-time limit, the main  orientation of the  domains will be
almost  completely parallel to  the shear  flow, and  therefore height
fluctuations  in  the surface  of  the  domains  grow in  a  direction
orthogonal  to $x$.   In $d=2$,  this implies  that the  only relevant
fluctuations  are in the  $y$, that  is $h$,  direction. On  the other
hand,  for $d= 3$,  there are  also fluctuations  growing in  the $\bf
x_\perp$  direction,  which  are   {\it  not}  described  by  equation
(\ref{burgers}).    These  two   cases  will   therefore   be  treated
separately.

\subsection{The case $d=2$}

In two dimensions the height  fluctuations of the surface are given by
the  fluctuations of the  field $h$.   Thus, as  a consequence  of the
scaling relation $h(x,t)=  b^\chi h'(x',t')$ (see Eq.\ (\ref{scale})),
the  height fluctuation  $\Delta$ grows  as  \beq \Delta  \sim h  \sim
t^{\chi/z} F(t/L_\parallel^z) \ ,  \eeq where the scaling function $F$
goes to a  constant for small argument and  $F(s)\sim s^{-\chi/z}$ for
$s\to\infty$. This means that if $t^{1/z} \ll L_\parallel$ the surface
grows   like  $\Delta\sim   t^{\chi/z}$,  whereas   if   $t^{1/z}  \gg
L_\parallel$,   we  have   $\Delta\sim   L_\parallel^\chi$.   We   can
incorporate  both  limits  in  the  form  \beq  \Delta\sim  \min\left(
t^{\chi/z}, L_\parallel^\chi \right) \ .
\label{chiodo}
\eeq In two dimensions we need  only consider models A ($\mu=1$) and B
($\mu=2$).

\subsubsection{Model A}

In this case the critical dimension is $d_c=5$, so for $d=2$ the shear
is  relevant.    From  the  former  sections  we   have  $\chi=0$  and
$z=1$. Equation (\ref{chiodo})  therefore implies that, whatever value
$L_\parallel$ takes, the height  fluctuation $\Delta$ will be of order
unity.   In \cite{orange}  it  has been  shown  that for  model A  the
transverse domain size  is $L_\perp \sim O(1)$. This  is an analytical
result   obtained   in  the   context   of  the   Ohta-Jasnow-Kawasaki
approximation.  This gives, \beq \Delta \sim L_\perp \quad \quad (d=2\
, \quad \quad  {\rm Model\ A}) \  .  \eeq We conclude that  model A in
two  dimensions  is  a  marginal  case,  and  we  cannot  exclude  the
possibility  that thermal  fluctuations  may in  this  case break  the
domains, giving rise to a stationary state.

\subsubsection{Model B}

For model B  we have $d_c=11/5>2$, and the  exponents are $\chi=-1/3$,
$z=4/3$. Also  in this  case, therefore,  we do not  need to  know the
coarsening   exponent   for   $L_\parallel$,   since   from   relation
(\ref{chiodo}) it is  clear that a negative value  of $\chi$ implies a
saturation  of $\Delta$  to a  constant value:  \beq \Delta  \sim O(1)
\quad \quad (d=2\  ,\quad \quad {\rm Model \ B})\  .  \eeq This result
opens up two different scenarios,  according to the the growth law for
$L_\perp$.  If  $L_\perp \sim t^{1/3}$,  as argued in  \cite{beppe} by
means  of numerical  experiments and  RG arguments,  then  $\Delta \ll
L_\perp$   and   the   domains   must  be   stable   against   thermal
fluctuations. If,  however, $L_\perp\sim  O(1)$, as suggested  by some
recent numerical simulations \cite{berthier},  then, as in model A, we
cannot  exclude the  possibility that  a  breaking of  the domains  by
thermal  fluctuations occurs.   Our result  shows that  a $L_\perp\sim
t^{1/3}$ growth  law and a  thermally induced stretching  and breaking
mechanism  are  not  compatible.  Conversely, if  a  thermally-induced
breaking of the domains  is clearly observed in numerical experiments,
this strongly suggests that the relation $L_\perp\sim O(1)$ holds.

\subsection{The case $d = 3$}

In three  dimensions the situation  is more complicated. First,  as in
$d=2$,   there  are   height  fluctuations   in  the   $y$  direction,
$\Delta_y\sim  h$, described  by equation  (\ref{burgers}).  Secondly,
there  are fluctuations  in the  $x_\perp$  direction, $\Delta_\perp$,
which  can  also  become  larger  than $L_\perp$,  and  that  are  not
described  by equation  (\ref{burgers}).  Thus,  before  assessing the
stability of the domains for $d = 3$ we must formulate an equation for
the description of these  latter fluctuations.  Fortunately, this will
turn  out  to be  a  linear equation,  such  that  no perturbative  RG
analysis is necessary.

In order to describe surface  fluctuations which grow in the $x_\perp$
direction  we have  to introduce  a new  height field  $h_\perp$ which
satisfies the  equation, \beq \partial_t h_\perp  +\gamma y \partial_x
h_\perp = {\cal L}h_\perp + \eta \ ,
\label{linear}
\eeq to  be compared  with (\ref{burgers}). The  operator $\cal  L$ is
still  given  at  low  momenta  by  ${\cal L}  \sim  \lambda  |{\bf  k
}|^{1+\mu}$.  Equation (\ref{linear}) is  linear, and therefore we can
work out the exponents exactly by means of simple scaling. By setting,
\beq x=bx',  \ \  y=b^{\zeta} y', \  \ h_\perp=b^{\chi} h_\perp',  \ \
t=b^{z}  t'  \  ,  \eeq  and imposing  scale  invariance  of  equation
(\ref{linear}),  we obtain  (with  the usual  hypothesis $\zeta  <1$),
\beqa    \gamma'   &=&    b^{z-1+\zeta}\,\gamma   \\    \lambda'   &=&
b^{z-\zeta(\mu+1)}  \lambda \\  D'  &=& b^{z-2\chi-\zeta\mu-1}  D \  ,
\eeqa  and  setting  to  zero  the scaling  dimensions  of  all  three
parameters      gives     \beq     z=\frac{\mu+1}{\mu+2},      \     \
\zeta=\frac{1}{\mu+2}, \  \ \chi=  - \frac{\mu+1}{2(\mu+2)} \  .  \eeq
Note that $\zeta$ is smaller than one, consistent with our assumption.
We see that $\chi$ is negative for all the three interesting values of
$\mu$  ($\mu=0,1,2$),  meaning  that  height  fluctuations  along  the
$x_\perp$ direction are always finite, $\Delta_\perp\sim O(1)$.

We have  to assess  now the physical  importance of $\Delta_y$  in the
context of domain coarsening. From the usual scaling relations we get,
\beq    \Delta_y\sim    h\sim    t^{\chi/z}    F(t/L_\parallel^z,    \
t/L_\perp^{z/\zeta}) \ .  \eeq In general, evaluating the magnitude of
$\Delta_y$ from this  relation is quite subtle, as  we need to compare
the  interfacial  coarsening  and  equilibrium  regimes  in  both  the
parallel and perpendicular directions.   However, as we discuss below,
in all cases of physical interest  we have $\chi \le 0$, implying that
the interfacial fluctuations saturate.

\subsubsection{Model A}

In the case $\mu=1$  and $d=3$, Eq.\ (\ref{nontrivial}), with $\mu=1$,
gives $\chi=-1/5$,  and therefore $\Delta_y\sim O(1)$. For  model A it
was  been found  in \cite{orange}  that $L_\perp\sim  t^{1/2}$, giving
\beq \Delta_y \ll L_\perp \quad \quad (d=3\ , \quad \quad {\rm Model \
A})\ .  \eeq In model  A, domains are therefore stable against thermal
fluctuations.

\subsubsection{Model B}

In   this  case   also   the  exponent   $\chi$   is  negative:   Eq.\
(\ref{nontrivial})  with  $\mu=2$   gives  $\chi=-2/7$,  and  $z=9/7$,
yielding \beq \Delta_y  \sim O(1) \quad \quad (d=3\  ,\quad \quad {\rm
Model \  B})\ .  \eeq Even  though no analytical  results or numerical
simulations studies are  available at the present time  for model B in
$d=3$, we certainly  expect $L_\perp$ to grow with  time in this case,
and therefore the domains to be stable.

\subsubsection{Model H}

For $\mu<1$, as we have  seen, we have a different critical dimension,
given by  Eq.\ (\ref{dc2}), which  is exactly three for  $\mu=0$. This
implies,  using  either   (\ref{free2})  or  (\ref{nontrivial}),  that
$\chi=0$ and $z=1$.  Once again,  this is the marginal case, with \beq
\Delta_y \sim O(1)  \quad \quad (d=3, \quad \quad {\rm  Model \ H})\ .
\eeq

\section{Summary}

Interfacial fluctuations  have been investigated  in systems subjected
to an external shear flow. Interfacial dynamics appropriate to systems
with  non-conserved scalar  order parameter  (``model  A''), conserved
scalar  order  parameter (``model  B''),  and  conserved scalar  order
parameter  coupled  to  hydrodynamic  flow  (``model  H'')  have  been
studied.   In each  case the  interfacial dynamics  is described  by a
similar equation, of the form  (\ref{burgers}), where $h$ is the local
height of  the interface and in  which the eigenvalue  spectrum of the
linear operator  $\cal{L}$ has the form  (\ref{spectrum}).  The models
differ principally in the numerical value of the exponent $\mu$, which
is given by 1,2 and 0 for models $A$, $B$ and $H$ respectively.

The  interface equations have  the form  of anisotropic  noisy Burgers
equations. In  each case,  exact renormalization group  (RG) arguments
determine the exponents $z$, $\zeta$, and $\chi$ that characterise the
coarsening,    anisotropy,   and    roughening   of    the   interface
respectively. In all cases, $\chi  \le 0$, implying that the thermally
induced  interfacial  width  approaches  a finite  limit  at  infinite
time. A consequence  of this result is that the  domain structure of a
coarsening system  under shear  is stable against  (sufficiently weak)
thermal fluctuations.

The general framework revealed by the exact RG relations was supported
by explicit  one-loop calculations  for $\mu \ge  1$.  For $\mu  < 1$,
however, no  one-loop equations consistent with  the expected critical
dimension, $d_c' =  3+2\mu$, could be derived. Whether  this is just a
technical  difficulty, or signals  some important  physical difference
between  the  regimes  $\mu \ge  1$  and  $\mu  < 1$,  merits  further
investigation.

\begin{center}
\begin{small}
{\bf ACKNOWLEDGEMENTS}
\end{small}
\end{center}

AC  thanks Antti  Kupiainen for  a  useful discussion.  This work  was
supported by EPSRC grant GR/L97698  (AJB and AC), and by Funda\c c\~ao
para a Ci\^encia e a Tecnologia grant BD/21760/99 (RDMT).

\end{multicols}


\begin{references} 

\bibitem{Review}  A.J. Bray, Adv.\  Phys.\ {\bf  43}, 357  (1994), and
references therein.

\bibitem{experiment} T.   Hashimoto, K.  Matsuzaka, E.   Moses, and A.
Onuki,  Phys.\  Rev.\  Lett.\  {\bf  74}, 126  (1995);  J.   L\"auger,
C. Laubner, and W. Gronski, Phys.\ Rev.\ Lett.\ {\bf 75}, 3576 (1995).

\bibitem{theory} T. Ohta, H. Nozaki, and M. Doi, J. Chem.\ Phys.\ {\bf
93}, 2664 (1991);  Y. N.  Wu, H. Skrdla, T. Lookman,  and S.  Y. Chen,
Physica A {\bf 239} (1-3), 428 (1997);
A. J. Wagner and J. M.  Yeomans, Phys.\ Rev.\ E {\bf 59}, 4366 (1999);
F. Corberi, G. Gonnella, and  A. Lamura, Phys.\ Rev.\ Lett.\ {\bf 81},
3852 (1998);  Phys.\ Rev.\ E {\bf  62}, 6621 (2000);  {\em ibid.} 8064
(2000); N.  P.  Rapapa and A.  J.  Bray, Phys.\ Rev.\ Lett.\ {\bf 83},
3856 (1999).

\bibitem{beppe} F.  Corberi, G. Gonnella, and A.  Lamura, Phys.\ Rev.\
Lett.\ {\bf 83}, 4057 (1999).

\bibitem{padilla} P.  Padilla and  S.  Toxvaerd, J. Chem.\ Phys.\ {\bf
106}, 2342 (1997).
 
\bibitem{orange} A. J. Bray and A.  Cavagna, J. Phys. A {\bf 33}, L305
(2000); A. Cavagna,  A. J. Bray, and  R. D. M. Travasso,  Phys. Rev. E
{\bf 62}, 4702 (2000).

\bibitem{shou} Z.   Y. Shou and  A. Chakrabarti, Phys.\ Rev.\  E, {\bf
61}, R2200 (2000).

\bibitem{berthier}
L. Berthier, Phys.\ Rev.\ E {\bf 63}, 051503 (2001).

\bibitem{burgers}
J. M. Burgers, {\it The nonlinear Diffusion Equation} (Reidel, Boston, 1974).

\bibitem{fns} 
D. Forster, D. R. Nelson, and M. J. Stephen, Phys. Rev. A {\bf 16},
732 (1977).

\bibitem{Hohenberg} P.C. Hohenberg, and B.I. Halperin, Rev. Mod. Phys.
{\bf 49}, 435 (1977).

\bibitem{ew} S. F. Edwards and D. R. Wilkinson, Proc.\ R.\ Soc.\ London, 
Ser.\ A {\bf 381}, 17 (1982). 

\bibitem{LTJZSO} J. S. Langer and L. A. Turski, Acta.\ Metall.\ {\bf 25}, 
1113 (1977); D. Jasnow and R. K. P. Zia, Phys.\ Rev.\ A {\bf 36}, 2243 
(1987); A. Shinozaki and Y. Oono, Phys.\ Rev.\ E {\bf 47}, 804 (1993).

\bibitem{shino} A. Shinozaki, Phys.\ Rev.\ E {\bf 48}, 1984 (1993). 

\bibitem{Bray98} A. J. Bray, Phys.\ Rev.\ E {\bf 58}, 1508 (1998). 

\bibitem{Siggia} E. D. Siggia, Phys.\ Rev.\ A {\bf 20}, 595 (1979). 

\bibitem{Furukawa} H. Furukawa, Phys.\ Rev.\ A {\bf 31}, 1103 (1985).

\bibitem{brief} 
A. J. Bray, A. Cavagna, and R. D. M. Travasso, Phys.\ Rev.\ E, in 
press. 

\bibitem{hwa}
T. Hwa and M. Kardar, Phys.\ Rev.\ A {\bf 45}, 7002 (1992). See also 
V. Becker and H. K. Janssen, Phys. Rev. E {\bf 50}, 1114 (1994).

\end{references}
\end{document}